\documentclass[journal=jacsat,manuscript=article]{achemso}

\usepackage{chemformula} 
\usepackage[T1]{fontenc} 
\usepackage{amsmath}
\usepackage{amssymb}
\usepackage{nomencl} 
\usepackage[normalem]{ulem}
\makenomenclature

\author{Michele Re Fiorentin}
\altaffiliation{These authors contributed equally to this work.}
\author{Francesca Risplendi}
\altaffiliation{These authors contributed equally to this work.}
\author{Clara Salvini}
\author{Juqin Zeng}
\author{Giancarlo Cicero}
\affiliation[Polito]
{Department of Applied Science and Technology, Politecnico di Torino,\\
corso Duca degli Abruzzi 24, 10129 Torino, Italy}
\author{Hannes J\'{o}nsson}
\email{hj@hi.is}
\affiliation[UIceland]
{Science Institute and Faculty of Physical Sciences, University of Iceland,\\ 107 Reykjavík, Iceland}
\alsoaffiliation[Brown]{Department of Chemistry, Brown University, Providence,\\ Rhode Island 02912, United States}

\title[An \textsf{achemso} demo]
{Silver electrodes are highly selective for CO in \ch{CO2} electroreduction due to interplay between voltage dependent kinetics and thermodynamics}

\abbreviations{IR,NMR,UV}
\keywords{American Chemical Society, \LaTeX}

\graphicspath{{./images/}}

\begin{document}


\begin{tocentry}
\includegraphics[height=4.45cm]{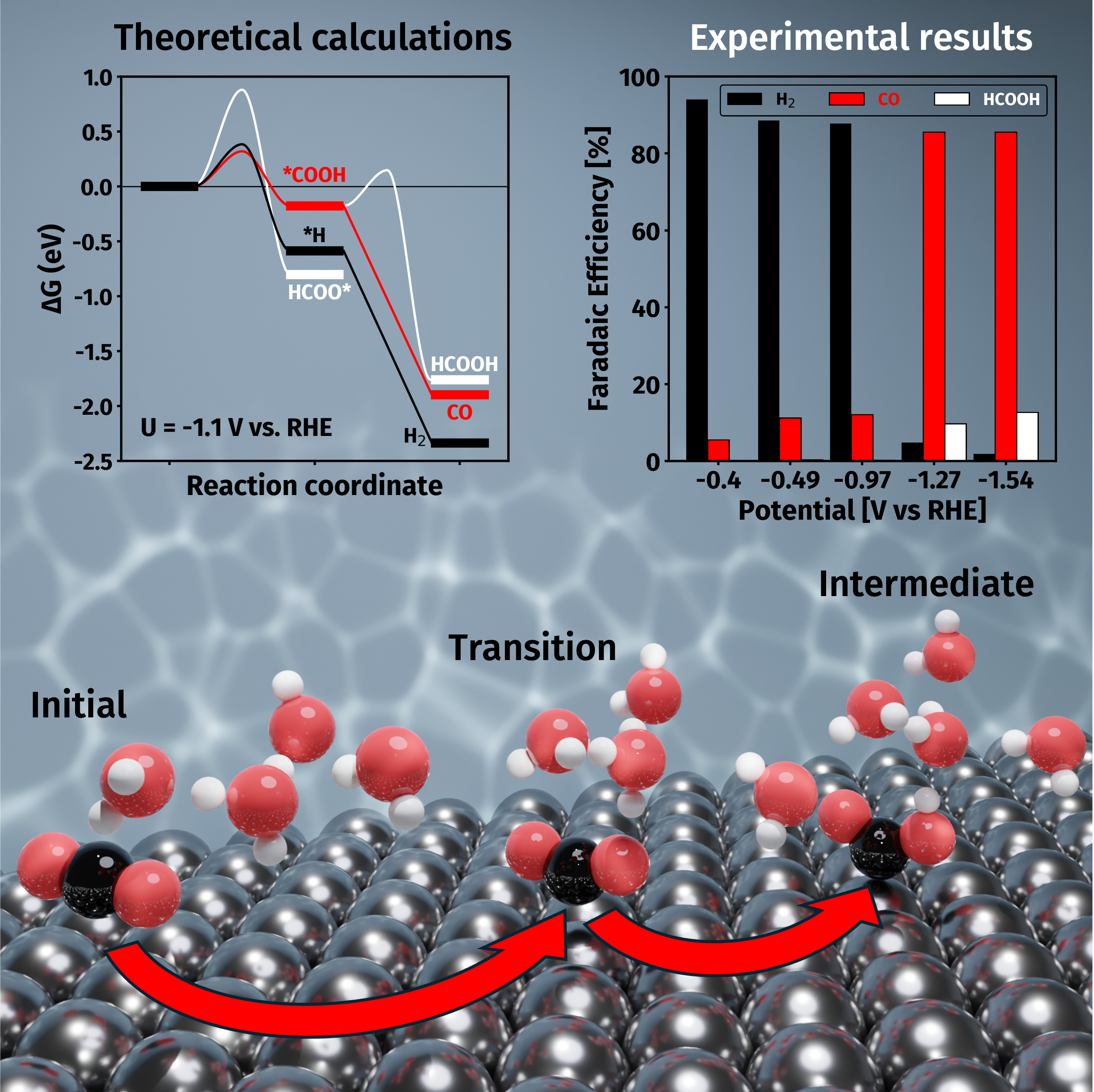}
\end{tocentry}

\begin{abstract}
Electrochemical reduction is a promising way to make use of \ch{CO2} as feedstock for generating renewable fuel and valuable chemicals. Several metals can be used in the electrocatalyst to generate CO and formic acid but hydrogen formation is an unwanted side reaction that can even be dominant. 
The lack of selectivity is in general a significant problem, but  
silver-based electrocatalysts have been shown to be highly selective, with faradaic efficiency of CO production exceeding 90\%, when the applied voltage is below $-1$ V {\it vs.} RHE.
In this voltage range,
only a small amount of hydrogen and formate is formed.
We present calculations of the activation free energy for the
various elementary steps as a function of applied voltage at the three low index facets, Ag(111), Ag(100) and Ag(110),
as well as experimental measurements on polycrystalline electrodes, to identify the reason for this high selectivity.
The formation of formic acid is suppressed,
even though it is thermodynamically favored,
because of the low coverage of adsorbed hydrogen and kinetic hindrance to the formation of the HCOO* intermediate,
while *COOH, a key intermediate in CO formation, is thermodynamically unstable until the applied voltage reaches $-1$ V {\it vs.} RHE, at which point the kinetics for its formation are more favorable than for hydrogen. 
The calculated results are consistent with experimental measurements carried out for acidic conditions and
provide an atomic scale insight into the high CO selectivity of silver-based electrocatalysts.
\end{abstract}
%
%
%

In the pursuit of carbon neutrality, a key strategy entails balancing emissions by capturing anthropogenic carbon dioxide and converting it into valuable products. One of the most promising transformative approach relies on the electrochemical reduction of carbon dioxide using engineered electrocatalysts capable of efficiently enhancing reaction kinetics, controlling reaction pathways, influencing product selectivity, and ensuring stability under working conditions\cite{MacDowell2017,Dibenedetto2014}. 
A wide range of electrocatalysts have been proposed so far for the \ch{CO2} reduction reaction (\ch{CO2}RR). \cite{Tomboc2020,Ma2021,Wang2020,Hu2018,Wang2024,Vavra2024,Zhang2024,Zhao2024} 
The choice of a specific catalyst determines the primary product of the reduction, which can vary from formate/formic acid or \ch{CO}, 
generated through a 2e$^-$ reduction, to multi-electron transfer products like alcohols and hydrocarbons.  
Additionally, the presence of competing reactions, such as the hydrogen evolution reaction (HER), significantly influences the catalyst selectivity. 
The complex  mechanisms involved in electrochemical reactions depend upon several factors \cite{Singh2015,Wagner2020} such as the morphology of the electrocatalyst \cite{Pan2020,Yu2020,Luc2017}, the composition of the electrolyte \cite{Pupo2019,Marcandalli2022,Sa2020}, its pH \cite{Varela2018,Wagner2020},
the \ch{CO2} partial pressure \cite{Song2020,Wagner2020}, the cell dimensions \cite{Tufa2020}, and, crucially, the applied potential.
The impact of these factors on \ch{CO2}RR can be assessed and predicted through theoretical modelling based on atomistic simulations. In particular, Density Functional Theory (DFT) provides valuable insight on the reaction mechanisms, enhancing our understanding of both the thermodynamics and kinetics of \ch{CO2}RR. 
The DFT thermochemical model (TCM)\cite{CHE} of electrocatalytic reactions has proven to be a robust method for predicting central thermodynamic quantities. It can provide key understanding and predictions of \ch{CO2}RR mechanisms that closely align with experimental observations and enable detailed examinations of the free energy landscape\cite{Montoya2017,Jonsson2017,Kulkarni2018}. However, by focusing solely on reaction thermodynamics, the TCM can only provide a lower bound on the reaction overpotential. More recently, efforts have been made to address the challenge of modelling and computing activation energies in electrochemical reactions. Several studies \cite{Singh2017,Jonsson2018,VandenBossche2019,VandenBossche2021,Govindarajan2022, Ma2022,Kastlunger2022,Karmodak2022} have investigated the kinetics of \ch{CO2}RR on metallic surfaces, offering additional insight and introducing a fundamental perspective on the problem that was previously missing.

In this Letter, we investigate the selectivity of silver surfaces towards \ch{CO2}RR at varying applied potential. We show that a combined theoretical study, encompassing both thermodynamics and kinetics, can fully clarify some crucial aspects of the behavior of silver electrocatalysts that still lack a fundamental explanation.
Silver-based electrocatalysts are widely employed in the electrocatalytic conversion of \ch{CO2} to \ch{CO}\cite{Dinh2018,VanDerHoek2024,Qin2023}, thanks to their remarkable selectivity. They achieve Faradaic efficiencies (FEs) for CO production exceeding 90\% and current densities suitable for industrial scale, reaching over 150 mA/cm$^2$\cite{Dinh2018,Fortunati2023,VanDerHoek2024,Qin2023,Monti2022}.
The production of HCOOH is observed in minimal quantities across varying applied potentials. Experimental evidence\cite{Hoshi1997,Hatsukade2014,Dutta2018,Clark2019} shows that the selectivity switches at intermediate overpotentials. Hydrogen evolution is favored at applied potentials $U\gtrsim-0.9$~V {\it vs.} RHE. As the bias is lowered, CO replaces \ch{H2} as the favored product, accompanied by an increase in the current density. A further change occurs at more cathodic potentials, when the FE of CO production drops significantly and the electrocatalyst selectivity switches back to HER. The analysis of the experimental current densities suggested that, unlike the former, this latter switch in selectivity can be attributed to mass transfer limitations\cite{Clark2019}. While well-known experimentally, the fundamental mechanisms behind this evident competition between \ch{CO} and \ch{H2} production at varying applied bias, as well as the extremely low formate production have not been extensively explored by atomistic modeling. In our DFT study of various \ch{CO2}RR pathways, along with the competing HER, we find that while \ch{HCOO}*, critical for formic acid production, is the most thermodynamically stable reaction intermediate, it is kinetically unfavorable across all applied biases. 
Crucially, we show that the selectivity crossover between CO and \ch{H2} production at intermediate potentials emerges only from a delicate interplay between reaction thermodynamics and kinetics.

Three crystal surfaces, namely (111), (100) and (110) were considered to model the silver electrode. We performed DFT electronic structure calculations with the VASP code\cite{VASP1,VASP2,VASP3,VASP4}, 
following the constant-potential computational setup established in \cite{VandenBossche2019,VandenBossche2021}.  
The Kohn-Sham equations were solved using the PAW\cite{Blochl1994} method and the RPBE functional\cite{Hammer1999}, with Monkhorst-Pack grids for Brillouin zone integrations\cite{Monkhorst1976,Pack1977}. Explicit \ch{H2O} molecules and the GLSSA13 implicit solvent model\cite{Gunceler_2013}, as implemented in the VASPsol\cite{VASPsol-Software,VASPsol2014-Dielectric,VASPsol2019-Electrolyte} plugin, were included to account for solvation effects. The implicit solvent model allows for the addition to the cell of a fractional number of electrons, compensated by the implicit counterion distribution. By varying the number of electrons in the DFT calculations it is possible to adjust the potential of the silver slab, obtained from its workfunction referenced to the bulk electrolyte. Standard conversions were then performed to obtain electrode potentials with respect to the RHE. 
The implicit solvent model assumes a bulk dielectric constant of water $\epsilon_\mathrm{bulk}=78.4$ and a Debye length of 3~\AA{}, corresponding to 1~M concentration of a monovalent symmetrical electrolyte. This is in line with the strong acidic conditions that we plan to study.
Transition states (TSs) were found as first-order saddle points along the minimum-energy path by means of nudged elastic band calculations followed by minimum-mode following\cite{Henkelman1999,Olsen2004,Gutierrez2017,Smidstrup2014} at constant potential. 
For further computational details, we refer the reader to the Supplementary Information.

\begin{figure}[t]
\centering
\includegraphics[width=0.5\textwidth]{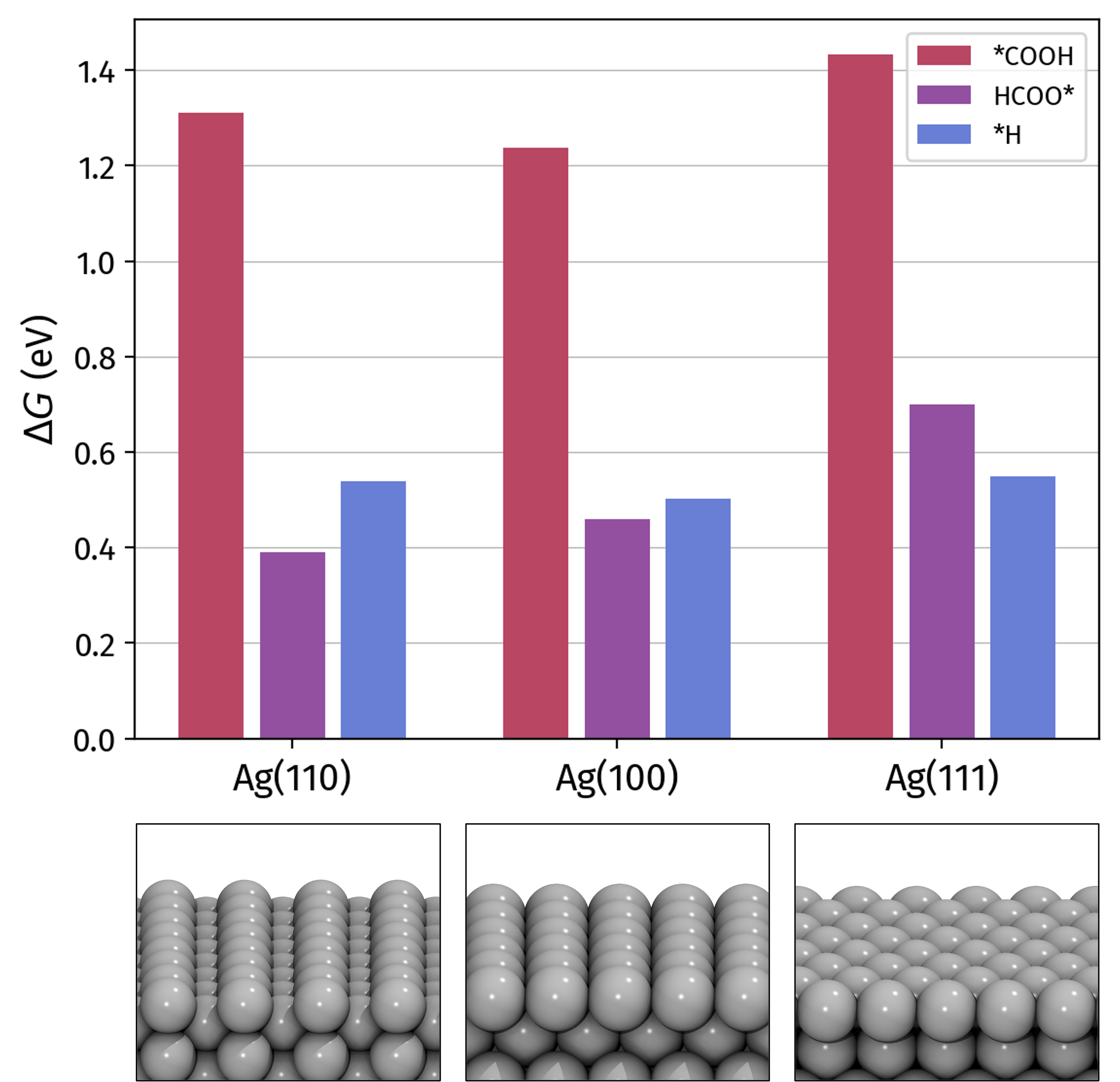}
\caption{Gibbs free energies of formation of *COOH (red bars), HCOO* (purple bars) and *H (blue bars) computed with the TCM. From left to right: Ag(110), Ag(100) and Ag(111) surfaces.}
\label{thermo}
\end{figure} 
A first thermodynamic study was performed to evaluate the Gibbs free energy of the first reaction steps of \ch{CO2RR} and HER via standard TCM. 
Specifically, we considered the two alternative proton-coupled electron transfer (PCET) reaction paths of \ch{CO2RR}, where \ch{CO2} is reduced to either CO through the *\ch{COOH} intermediate
\begin{align}
\label{eq:CO2_H+e->*COOH}
    *\,+\,&\rm{CO}_2
    \xrightarrow{\rm{H}^++\rm{e}^-}
    {}^*\rm{COOH}\\
\label{eq:*COOH->CO}
    &{}^*\rm{COOH}\xrightarrow{\rm{H}^++\rm{e}^-}\rm{CO}+\rm{H}_2\rm{O},
\end{align}
or to HCOOH through the \ch{HCOO}* intermediate
\begin{align}
\label{eq:CO2_H+e->HCOO*}
    *\,+\,&\rm{CO}_2\xrightarrow{\rm{H}^++\rm{e}^-} \rm{HCOO}^*\\
\label{eq:*HCOO->HCOOH}
    &\rm{HCOO}^*\xrightarrow{\rm{H}^++\rm{e}^-}\rm{HCOOH}.
\end{align}
From Figure \ref{thermo}, reporting the Gibbs free energies of formation calculated on the three silver surfaces, we can notice that the formation of *\ch{COOH}, Eq.~(\ref{eq:CO2_H+e->*COOH}), is thermodynamically less favored than that of \ch{HCOO}*, Eq.~(\ref{eq:CO2_H+e->HCOO*}), on all silver surfaces.
Interestingly, 
the Gibbs free energy of \ch{HCOO}* formation increases in the order Ag(110) > Ag(100) > Ag(111), as the degree of surface packing
increases and the atoms at the surface are less undercoordinated. 
Finally, we also considered HER, through the successive Volmer and Heyrovsky steps, as
\begin{align}
\label{eq:H+e->*H}
    *\,&\xrightarrow{\rm{H}^++\rm{e}^-}{}^*\rm{H}\\
\label{eq:*H->H2}
    &{}^*\rm{H}\xrightarrow{\rm{H}^++\rm{e}^-}\rm{H}_2.
\end{align}
Figure \ref{thermo} shows that the formation of *\ch{H} is disfavored compared to \ch{HCOO}* on all surfaces, except for Ag(111), 
where hydrogen is adsorbed on the fcc site with higher stability. The results of this preliminary analysis confirm previous theoretical findings: based on reaction thermodynamics, silver electrodes should be expected to produce \ch{HCOOH} and \ch{H2}, contrary to what is observed experimentally
\cite{Hoshi1997,Rosen2015,Clark2019,Yoo2016}. 
Clearly, the TCM alone cannot reliably predict the selectivity of Ag surfaces and a more comprehensive investigation, incorporating explicitly applied potentials and TS searches, is necessary.
\begin{figure}[tb]
\centering
\includegraphics[width=.8\linewidth]{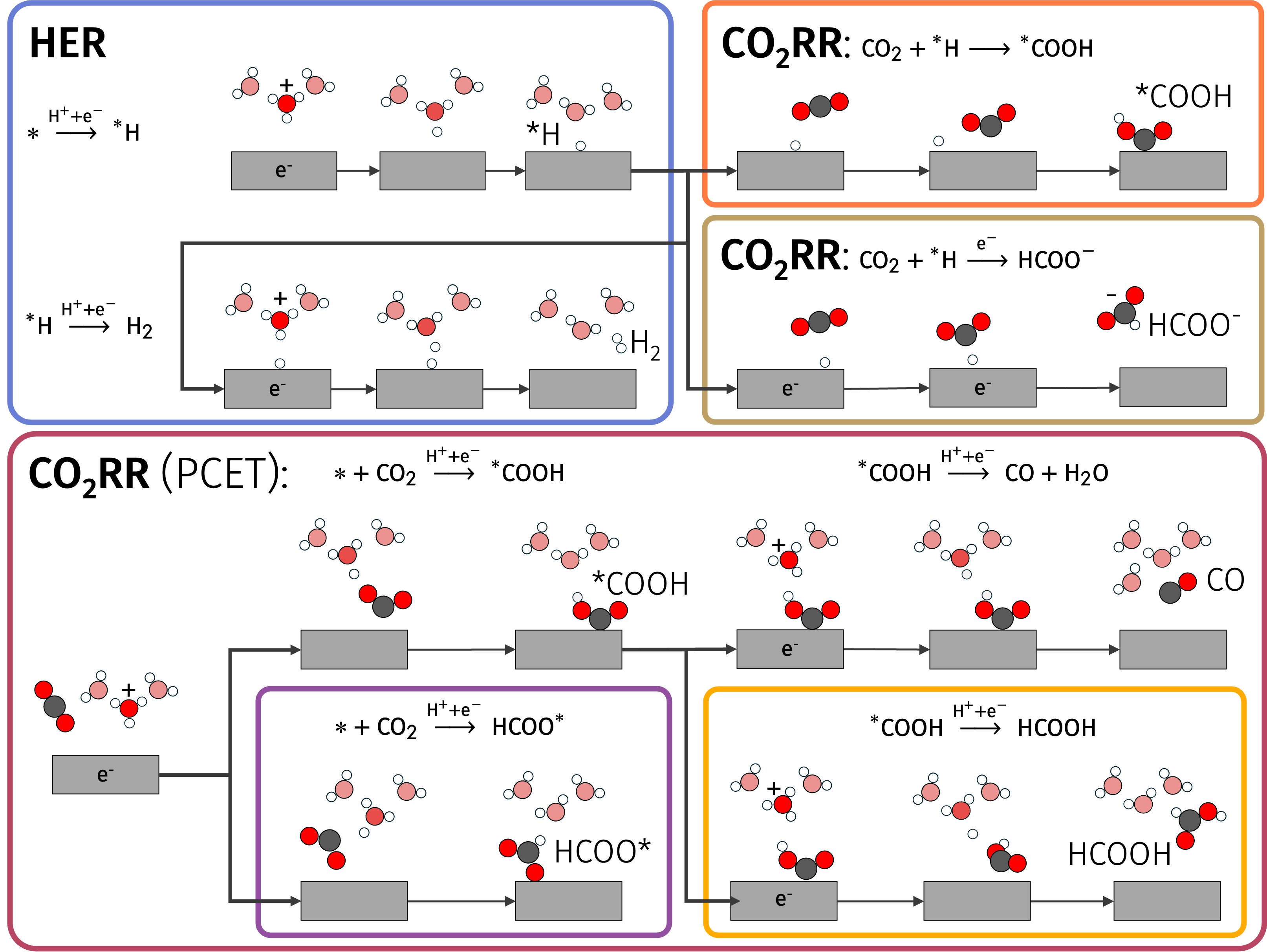}
\caption{Diagram of the studied reaction pathways, TSs and intermediate states for \ch{CO2RR} and HER. Carbon, oxygen and hydrogen atoms are represented by black, red and white circles respectively. Grey rectangles represent the silver slab.}
\label{fig:scheme}
\end{figure}

We identified the TSs relevant to \ch{CO2RR} and HER under constant external potential, obtaining the corresponding activation grand-canonical energy\cite{VandenBossche2019}, $\Delta \Omega^\ddagger$. 
In this study, we focus on the kinetics of reactions occuring in acidic conditions, where the proton donors are solvated hydronium ions, \ch{H3O+}. 
We also assume efficient mass transport, ensuring that the hydronium concentration remains sufficiently high at the interface to suppress reactions involving water as the proton donor. This condition is indeed achieved in our experimental setup (see Supplementary Information).
\begin{figure}[t]
    \centering
    \includegraphics[width=\linewidth]{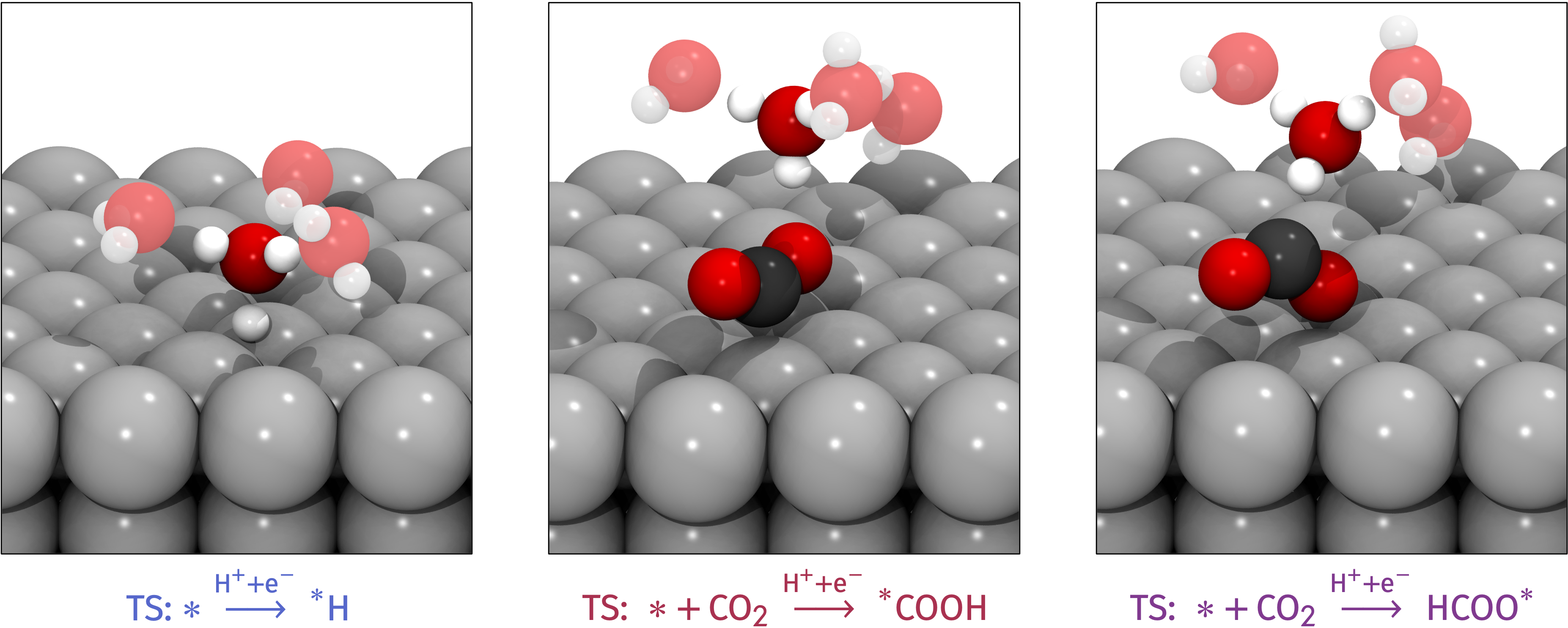}
\caption{Geometries of three representative TSs on Ag(111) at $U=-0.8$~V {\it vs.} RHE. Left panel: HER step in Eq.~(\ref{eq:H+e->*H}). Central panel: \ch{CO2RR} step in Eq.~(\ref{eq:CO2_H+e->*COOH}). Right panel: \ch{CO2RR} step in Eq.~(\ref{eq:CO2_H+e->HCOO*}). Carbon, oxygen, hydrogen and silver atoms are represented by black, red, white and gray spheres, respectively. Lighter colors are used to mark spectator water molecules.}
    \label{fig:TSgeom}
\end{figure}
The studied reaction pathways, TSs and intermediate states are reported in the scheme in Figure~\ref{fig:scheme}.
Beside the PCET reaction in Eq.~(\ref{eq:CO2_H+e->*COOH}), the *\ch{COOH} intermediate can be obtained from a \ch{CO2} molecule and a surface-adsorbed hydrogen atom, as
\begin{align}
\label{eq:CO2_*H->*COOH}
    \rm{CO}_2\,+\,{}^*\rm{H}\longrightarrow{}^*\rm{COOH}.  
\end{align}
Similarly, \ch{CO2} can be also directly transformed into \ch{HCOO^-} with an adsorbed *H and a transferred electron, as
\begin{align}
\label{eq:CO2_e->HCOO-}
    \rm{CO}_2\,+\,{}^*\rm{H}\overset{\rm{e}^-}{\longrightarrow}\rm{HCOO}^-.  
\end{align}
\ch{HCOO^-} then readily transforms into \ch{HCOOH} in a strongly acidic environment. Finally, formic acid can also be obtained from the *\ch{COOH} intermediate as
\begin{align}
\label{eq:*COOH->HCOOH}
    {}^*\rm{COOH}\xrightarrow{\rm{H}^++\rm{e}^-}\rm{HCOOH},
\end{align}

\begin{figure}[t]
    \centering
    \includegraphics[width=\linewidth]{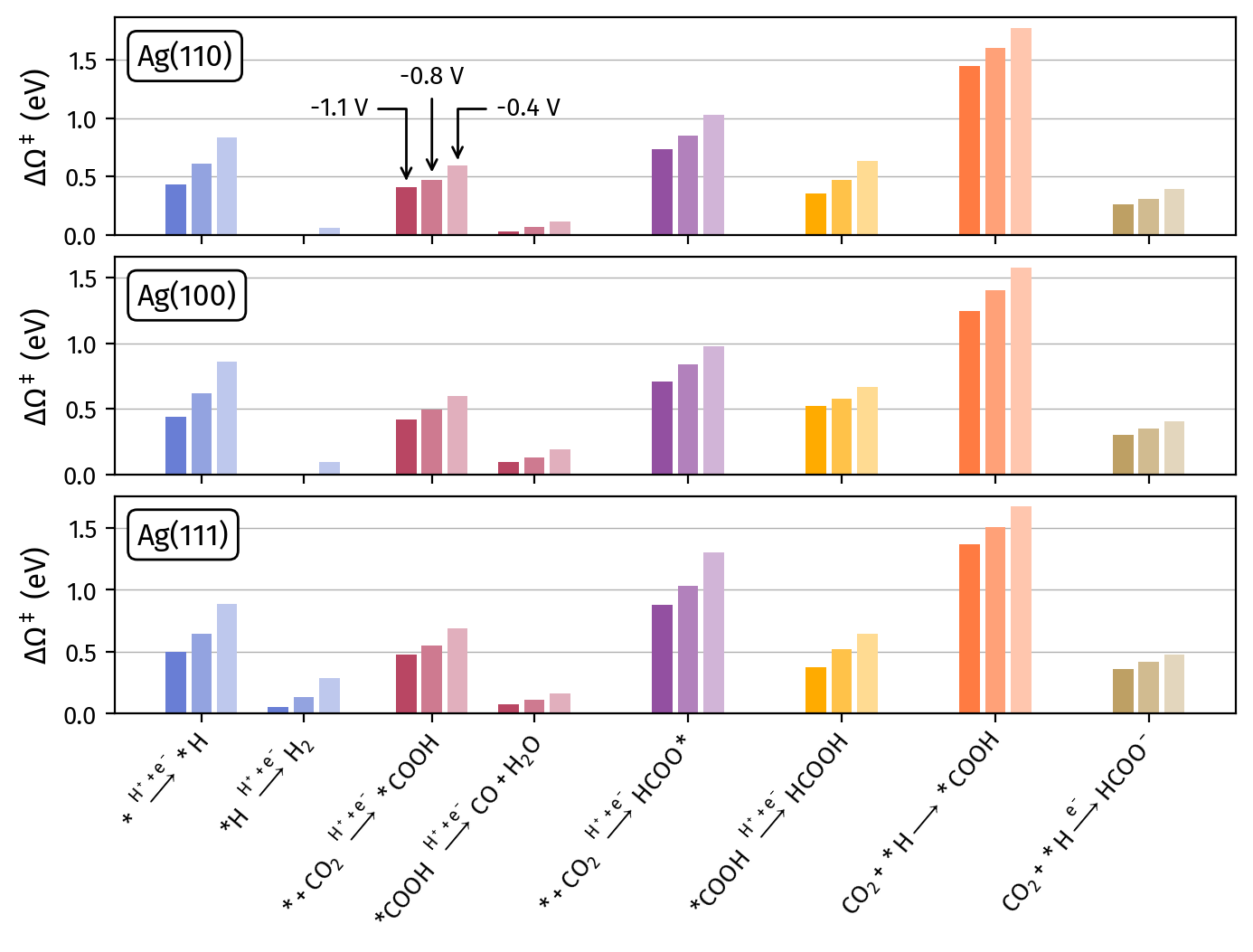}
    \caption{Grand-canonical activation energies $\Delta\Omega^\ddagger$ of the studied steps of HER and \ch{CO2RR}. Top to bottom: Ag(110), Ag(100) and Ag(111) surfaces. Each group represents a reaction step and consists of three bars, corresponding to the three studied potentials, arranged from left to right (darker to lighter shades): $U=-1.1,\,-0.8,\,-0.4$~V~{\it vs.}~RHE.}
    \label{fig:omega_comparison}
\end{figure}
The geometries of three representative TSs on Ag(111) at $U=-0.8$~V {\it vs.} RHE are shown in Figure~\ref{fig:TSgeom}. The geometries of the other investigated TSs are reported in the Supporting Information. The $\Delta\Omega^\ddagger$s at the three studied potentials, $U=-0.4,\,-0.8,\,-1.1$~V {\it vs.} RHE, are reported in Figure~\ref{fig:omega_comparison}.
Comparing the activation grand-canonical energies, it appears that for \ch{CO2RR} the formation of \ch{HCOO-} via reaction mechanism (\ref{eq:CO2_e->HCOO-}) enjoys the lowest kinetic barriers on all investigated silver surfaces. On the contrary, the PCET mechanism favors the formation of the *\ch{COOH} intermediate, while the activation energy for the formation of \ch{HCOO}* remains higher than 0.5 eV on all surfaces at all studied potentials. 
Consequently, considering PCET mechanisms, \ch{CO2RR} tends to proceed through the *\ch{COOH} intermediate, Eq.~(\ref{eq:CO2_H+e->*COOH}), followed by \ch{CO} production, Eq.~(\ref{eq:*COOH->CO}), which has much lower barriers than the step in Eq.~(\ref{eq:*COOH->HCOOH}), forming \ch{HCOOH}. Finally, the mechanism in Eq.~(\ref{eq:CO2_*H->*COOH}) exhibits the highest activation energies on all silver surfaces, indicating that this reaction pathway is highly disfavored.
Considering HER, the Volmer step in Eq.~(\ref{eq:H+e->*H}) exhibits a decreasing activation barrier from approximately 1.0 eV at $-0.4$~V {\it vs.} RHE to around 0.4 eV when the cathodic potential is lowered to $-1.1$~V {\it vs.} RHE. Once hydrogen is adsorbed, the Heyrovsky reaction in Eq.~(\ref{eq:*H->H2}) proceeds with very low activation energies across the entire potential range, becoming barrierless on Ag(100) and Ag(111) at potentials lower than about $-0.8$~V {\it vs.} RHE. The Volmer step of HER is therefore quickly followed by the Heyrovsky step, producing \ch{H2}, rather than the \ch{CO2RR} step in Eq.~(\ref{eq:CO2_e->HCOO-}), where the adsorbed hydrogen from the Volmer step reacts with \ch{CO2} to form \ch{HCOO-}.
Globally then, the suppressed production of formate on Ag can be attributed to two main factors. First, the high kinetic barrier for \ch{HCOO}* formation via the PCET in Eq.~(\ref{eq:CO2_H+e->HCOO*}). Second, the very low *\ch{H} coverage \cite{Jonsson2018}, resulting from high hydrogen adsorption energy \cite{} and low Heyrovsky activation energy, which hinders the direct \ch{HCOO-} formation via Eq.~(\ref{eq:CO2_e->HCOO-}).
From the activation grand-canonical energies in Figure~\ref{fig:omega_comparison} we can then conclude that on silver surfaces the \ch{CO2RR} proceeds to CO along the PCET steps in Eqs.~(\ref{eq:CO2_H+e->*COOH}), (\ref{eq:*COOH->CO}), passing through the *COOH intermediate. In competition, silver also produces hydrogen through the Volmer-Heyrovsky mechanism. All other pathways appear to be comparatively disfavored.
From the analysis of $\Delta\Omega^\ddagger$ of HER and \ch{CO2RR}, we may conclude that the latter is the most kinetically favored reduction reaction on all silver surfaces, across the whole potential window. 
However, this finding contradicts the experimental evidence mentioned in the introduction, which indicates a predominant HER at low overpotential.
\begin{figure}[tb]
\centering
\includegraphics[width=\linewidth]{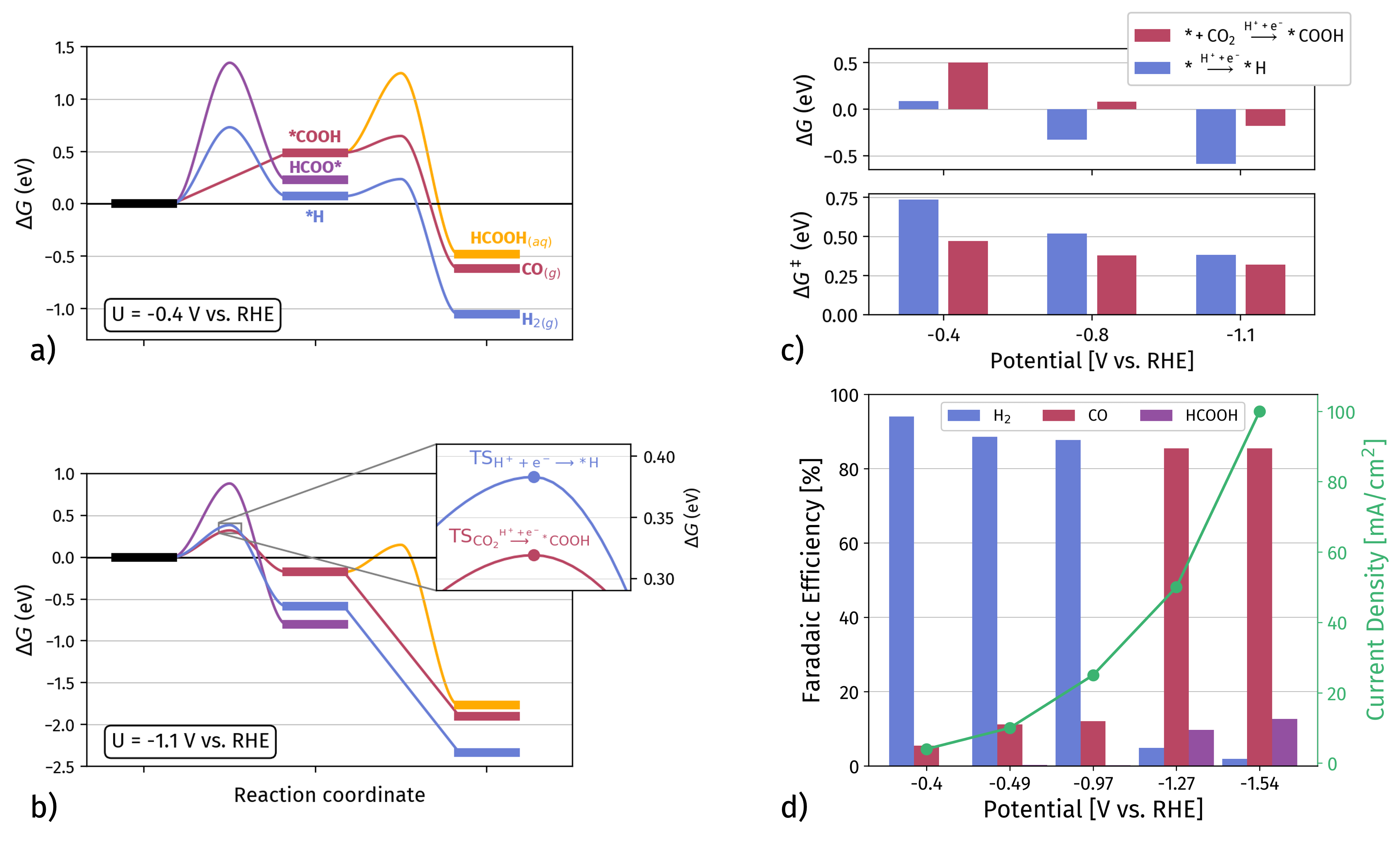}
\caption{(a) and (b): Grand-canonical free energy variations $\Delta G$ on Ag(111) along \ch{CO2RR} and HER at $U=-0.4$~V {\it vs.} RHE (a) and $U=-1.1$~V {\it vs.} RHE (b). The inset in panel (b) shows a zoom on the activation grand-canonical free energies of *\ch{H} production in HER (blue line) and *\ch{COOH} formation in \ch{CO2RR}. (c) Comparison of reaction (upper panel) and activation (lower panel) grand canonical free energies for reaction steps in Eqs.~(\ref{eq:CO2_H+e->*COOH}) (red bars) and (\ref{eq:H+e->*H}) (blue bars). (d) Experimental FEs of \ch{CO2RR} to CO (red bars) or to HCOOH (purple bars) and HER (blue bars) of a synthesized polycristalline Ag electrocatalyst in a pH 2 electrolyte. }
\label{fig:Thermo_Kinetics}
\end{figure}

To obtain a complete and comprehensive description of the reaction mechanisms, it is then essential to consider both thermodynamics and kinetics together. Due to the similar thermodynamic and kinetic results, we focus on the Ag(111) surface. In Figure~\ref{fig:Thermo_Kinetics}a,b we report the variation in grand-canonical free energy, $\Delta G$, along HER and \ch{CO2RR}, including intermediate and transition states. Reaction grand-canonical free energies are computed at constant applied potential, employing the described hybrid explicit-implicit solvation setup. Activation grand-canonical free energies, $\Delta G^\ddagger$, are derived from the grand-canonical energies of the transition states found with constant-potential calculations. 
In the calculation of grand-canonical free energies we consider hydronium ions in equilibrium in bulk water.
Grand-canonical free energies and activation free energies at $U=-0.4,\,-0.8\,-1.1$~V {\it vs.} RHE are also reported in Figure~\ref{fig:Thermo_Kinetics}c.

At low overpotential, $U=-0.4$~V {\it vs.} RHE, as shown in Figure~\ref{fig:Thermo_Kinetics}a, the states with the highest grand-canonical free energies, when considering full $\Delta G$ variations along the \ch{CO2RR} pathways, are the TSs of \ch{HCOO}* formation, Eq.~(\ref{eq:CO2_H+e->HCOO*}), purple line, and \ch{HCOOH} formation from *\ch{COOH}, Eq.~(\ref{eq:*COOH->HCOOH}), yellow line. Interestingly, for the *\ch{COOH} formation, Eq.~(\ref{eq:CO2_H+e->*COOH}), red line, it is the intermediate *\ch{COOH} that shows the highest grand-canonical free energy. This can also be seen in Figure~\ref{fig:Thermo_Kinetics}c. Here, at $U=-0.4$~V {\it vs.} RHE, the activation grand-canonical free energy $\Delta G^\ddagger$ of *\ch{COOH} production (red bar in the lower panel), computed from the grand-canonical electronic energy of the TS, is lower than the reaction grand-canonical free energy $\Delta G$ (red bar, upper panel). Hence, no actual kinetic barrier appear along the grand-canonical free energy reaction path in Figure~\ref{fig:Thermo_Kinetics}a. 
In this situation, the backward step, \ch{CO2}$\,\longleftarrow\,$*\ch{COOH}, is then both thermodynamically favored and kinetically barrierless.
Consequently, at low overpotential, \ch{CO2RR} is overall suppressed: the production of *\ch{COOH} is hindered by the unfavorable thermodynamics, while the formation of \ch{HCOO}* is prevented by the large kinetic barrier.
In this range of applied voltage, HER dominates, even though the $\Delta G^\ddagger$ of the Volmer step remains above 0.5~eV, in line with the value of $\Delta\Omega^\ddagger$ in Fig.~\ref{fig:omega_comparison}. 
As the bias is lowered to $U=-1.1$~V {\it vs.} RHE, Figure~\ref{fig:Thermo_Kinetics}b, the intermediate state *\ch{COOH} is stabilized, its grand-canonical free energy of formation becomes lower and a kinetic barrier emerges, see also Figure~\ref{fig:Thermo_Kinetics}c. 
*\ch{COOH} can then be formed and further reduced to \ch{CO} without kinetic barrier. At the same time, \ch{HCOOH} formation from *\ch{COOH} still shows a nonzero activation energy, making this step less favorable. Most importantly, as highlighted in the inset of Figure~\ref{fig:Thermo_Kinetics}b, the kinetic barrier for *\ch{COOH} formation is lower than that of the Volmer step of HER, resulting in a more favorable CO production compared to \ch{H2}. Therefore, at intermediate potentials the selectivity switches from HER to \ch{CO2RR}, specifically with almost only CO production.

To validate our theoretical investigations, focusing specifically on reaction steps involving hydronium ions as proton donors, we carried out electrochemical characterizations of silver electrodes in an acidic environment. Indeed, most \ch{CO2RR} experiments in the literature were performed at neutral or alkaline pH, and the acidic reaction has received minor attention\cite{Yu_ChemCat_2023,Zhang2023}. A polycrystalline silver electrode was prepared via sputtering\cite{Monti2022}, and tested in flow cell with an electrolyte at pH 2. More details about the experimental procedure are reported in the Supporting Information.
As shown in Figure~\ref{fig:Thermo_Kinetics}c, CO and \ch{H2} emerge as the primary gaseous products, with small amounts of HCOOH. The expected trend is clearly observable: a shift from HER, favored up to intermediate potentials of about $-1$~V {\it vs.} RHE, to CO production at more negative potentials, accompanied by an increase in the current density. The cumulative FE values for CO, \ch{H2}, and HCOOH approximate 100\%, suggesting minimal formation of other products. 
These results are in good agreement with our proposed theoretical model which predicts a selectivity switch at around the same applied potential values.

We conclude that, in critical cases such as the reactions investigated on silver electrodes, simpler or partial approaches may fail to provide a correct and complete understanding of the reaction mechanisms. In these situations, only by integrating the theoretical estimation of both reaction thermodynamics and kinetics we can obtain a detailed and satisfactory explanation of the observed selectivity in \ch{CO2RR}. Our focus on the case study of Ag electrocatalysts demonstrates that comprehensive DFT modeling is essential to reliably understand the underlying mechanisms of electrochemical reactions and derive meaningful predictions.

\begin{acknowledgement}
MRF, FR and GC acknowledge the High-Performance Computing, Big Data, and Quantum Computing Research Centre, established under the Italian National Recovery and Resilience Plan (PNRR).
\end{acknowledgement}

\begin{suppinfo}
The supporting information accompanying this publication provides additional details and 
data essential for understanding and reproducing the results presented in the main 
manuscript. 
Below is a comprehensive list of the contents included in the supporting information file ``supporting$\_$info.pdf'':
\begin{itemize}
      \item Computational Details: DFT setup and  Thermodynamics and Kinetics.
      \item Detailed Explanation of Reaction Mechanisms.
      \item Geometries of Transition States.
\end{itemize}
    
\end{suppinfo}

\bibliography{paper}

\end{document}